\documentclass[11pt,twocolumn]{article}

\usepackage[utf8]{inputenc}
\usepackage[T1]{fontenc}
\usepackage{mathptmx}
\usepackage[margin=2cm]{geometry}
\usepackage{graphicx}
\usepackage{amsmath}
\usepackage{natbib}
\usepackage{hyperref}
\usepackage{xcolor}
\usepackage{booktabs}
\usepackage{caption}

\hypersetup{colorlinks=true,citecolor=blue!60!black,linkcolor=blue!60!black,urlcolor=blue!60!black}

\title{Multidimensional Profiles of Critical Thinking in Physics Labs:\\Latent Structure, Instructional Change, and Connections to Physics Identity}

\author{Marcus Kubsch$^{1}$, Natasha G.\ Holmes$^{2}$, Antti Lehtinen$^{3}$\\[6pt]
{\normalsize\itshape $^{1}$Department of Physics, Freie Universit\"at Berlin, Germany}\\
{\normalsize\itshape $^{2}$Department of Physics, Cornell University, USA}\\
{\normalsize\itshape $^{3}$Department of Physics, University of Jyv\"askyl\"a, Finland}}

\date{}

\begin{document}
\twocolumn[
  \begin{@twocolumnfalse}
  \maketitle
  \begin{abstract}
The Physics Lab Inventory of Critical Thinking (PLIC) measures three components of students' critical thinking in physics labs: evaluating data, evaluating methods, and proposing next steps. Prior work has analyzed these components in isolation or as a composite score. In this study, we apply latent profile analysis (LPA) to the three PLIC scales using a large, multi-institutional dataset of 5,513 matched pre/post student records to identify characteristic response patterns across the three components simultaneously. At both pre- and post-instruction, a two-profile solution best fit the data. Profile composition shifted substantially over instruction, with 48.4\% of students in the lower-performing profile at pre-test transitioning to the higher-performing profile at post-test, while 43.6\% of students moved in the opposite direction. Course type was statistically associated with profile membership at both timepoints, though the effect was small (Cram\'er's $V \approx 0.10$). To examine the relationship between profile transitions and students' affective development, we estimated cross-lagged panel models (CLPMs) linking profile membership to belonging, recognition, self-efficacy, and agency. Belonging emerged as the principal upstream predictor, prospectively predicting recognition, self-efficacy, agency, and higher-knowledge profile membership. Agency and self-efficacy formed a reciprocal but asymmetric loop, with the path from agency to later self-efficacy being stronger. Recognition functioned primarily as a downstream construct over this timescale. These results provide the first person-centered, multidimensional characterization of PLIC performance and demonstrate that epistemic and identity-related constructs are interlinked in physics lab learning.
  \end{abstract}
  \vspace{1em}
  \end{@twocolumnfalse}
]

\section{Introduction}

Physics laboratory courses occupy a distinctive position in undergraduate education: they are among the most resource-intensive components of the curriculum, and they offer students one of their few structured opportunities to engage directly with experimental reasoning \citep{HolmesWieman2018, AAPT2014}. Over the past decade, a growing body of evidence has challenged the traditional use of labs primarily to reinforce lecture content, showing that such approaches produce little measurable learning gain \citep{Holmes2017, WiemanHolmes2015}. In response, professional organizations and researchers have called for labs to focus instead on developing students' critical thinking and experimentation skills \citep{AAPT2014, Holmes2015}.

This shift in goals has created a corresponding need for assessment instruments capable of measuring the targeted outcomes. The Physics Lab Inventory of Critical Thinking (PLIC) was developed to fill this role \citep{Walsh2019}. The PLIC measures students' ability to evaluate experimental data, evaluate methods, and propose productive next steps for an investigation. Since its validation, the PLIC has been administered at over 100 institutions and has accumulated a dataset of more than 30,000 student responses \citep{Walsh2022}. Research with the PLIC has demonstrated that skills-focused lab instruction produces measurable improvements in critical thinking compared to concepts-focused instruction, and that these effects are equitable across demographic groups \citep{Walsh2022, Smith2020}.

Despite these advances, most analyses of PLIC data have relied on a single composite score, treating students' critical thinking as a unidimensional quantity. The three-factor structure of the PLIC---evaluating data, evaluating methods, and suggesting next steps---has been validated \citep{Walsh2019} and confirmed in cross-cultural contexts \citep{Pirinen2023}, yet only one study has used these subscales to examine students' competence profiles in a fine-grained way \citep{LehtKub2026}. That analysis, conducted with a modest Finnish sample, identified three distinct profiles of incoming students, suggesting that students' critical thinking is not simply higher or lower but differs in configuration across the three components. What remains unknown is whether similar profile structures appear in a large, multi-institutional, international dataset and how those profiles change over the course of instruction.

A second open question concerns the relationship between students' epistemic performance on the PLIC and their affective experiences in the lab. Research on physics identity---drawing on frameworks from \citet{Hazari2010} and \citet{CarloneJohnson2007}---has established that constructs such as belonging, recognition, self-efficacy, and agency shape students' persistence and engagement in physics. \citet{Kalender2021} demonstrated that these constructs relate to students' experiences specifically in lab settings, and \citet{Stump2023} showed that students' lab roles influence their identity development. However, the relationship between identity-related constructs and the \emph{epistemic quality} of students' lab engagement---as measured by instruments like the PLIC---has not been examined longitudinally. Understanding this interplay matters because belonging, self-efficacy, and agency may serve as preconditions for productive critical thinking, or they may develop as a consequence of it.

In this study, we address two research questions:
\begin{enumerate}
    \item What latent profiles of critical thinking emerge across the three PLIC scales in a large, multi-institutional dataset, and how do students transition between profiles over a course of instruction?
    \item What is the longitudinal relationship between students' PLIC profile membership and their physics identity constructs (belonging, recognition, self-efficacy, and agency)?
\end{enumerate}

\section{Background}

\subsection{Critical Thinking in Physics Labs}

A central goal of science education is to teach students to think critically about data and models \citep{Holmes2015}. In a physics lab context, critical thinking involves making decisions about what to trust and what to do: evaluating whether data support a model, comparing the quality of different experimental methods, and reasoning about productive next steps for an investigation \citep{Walsh2019}. \citet{Holmes2015} demonstrated that structuring lab instruction around iterative comparison cycles---where students repeatedly compare data to models and decide how to improve their measurements---produced dramatic and lasting gains in students' critical thinking behaviors compared to traditional verification labs.

The PLIC was developed to assess these skills at scale \citep{Walsh2019, Quinn2018}. The instrument presents respondents with a scenario in which two groups of physicists investigate a mass-on-a-spring system using different experimental designs. Students answer questions evaluating each group's data (the ``evaluate data'' component), comparing the groups' methods (the ``methods'' component), and proposing what each group should do next (the ``next steps'' component). Scoring is based on expert physicist responses and uses a weighted scheme that rewards selection of high-value response options.

Large-scale analyses with the PLIC have examined the effects of lab type on student outcomes. \citet{Walsh2022} analyzed data from over 20,000 students across more than 100 institutions and found that labs focused on developing experimentation skills produced gains in both critical thinking (measured by the PLIC) and views about experimental physics (measured by the E-CLASS), while concepts-focused labs did not. Pedagogical features associated with these gains included activities involving decision making and communication, while modeling activities had smaller effects. \citet{Smith2020} used a controlled quasi-experimental design to show that structured lab instruction centered on iterative experimentation improved students' critical thinking as measured by the PLIC.

\subsection{Person-Centered Approaches to Assessment}

Variable-centered analyses---comparing group means on a composite score or examining individual item performance---are the dominant approach in assessment research, including PLIC research to date. These analyses assume that the population is homogeneous with respect to the relationships among measured variables. Person-centered approaches, by contrast, identify subpopulations of individuals who share similar configurations of scores across multiple dimensions \citep{Lanza2013}. Latent profile analysis (LPA) is one such method: it models a population as a finite mixture of unobserved subgroups, each characterized by its own set of means across observed variables.

Person-centered methods are theoretically appropriate when constructs are multidimensional and when individuals may differ not just in overall level but in the \emph{pattern} of their scores. The three-factor structure of the PLIC makes it a natural candidate for this approach. A student who scores high on evaluating data but low on suggesting next steps represents a qualitatively different configuration of critical thinking than a student with the reverse pattern. Composite scores obscure these configurations.

In physics education research, person-centered methods have seen growing use. Latent class and profile analyses have been applied to characterize response patterns on attitude surveys \citep{Douglas2014} and to identify subgroups of students with distinct configurations of beliefs about experimental physics. \citet{LehtKub2026} applied LPA to PLIC subscale data from Finnish physics students and identified three profiles: one characterized by strong data evaluation, one by relatively uniform performance, and one by low scores across all components. That study, however, was limited by a small sample ($N = 174$) drawn from a single institution, and it examined only incoming students without a post-instruction comparison.

\subsection{Physics Identity and Lab Learning}

Physics identity---the extent to which a person sees herself as a ``physics person''---has been shown to predict persistence and career choice in physics \citep{Hazari2010}. Building on earlier work by \citet{CarloneJohnson2007}, \citet{Hazari2010} proposed a framework in which physics identity is shaped by interest, recognition (being seen as a physics person by others), and competence/performance beliefs. Subsequent research has highlighted the role of belonging \citep{Rainey2018, Lewis2017}, self-efficacy \citep{Kalender2020, Bandura1997}, and agency \citep{Kalender2021} in students' physics experiences, particularly in lab contexts.

Lab courses provide a distinctive setting for identity development because they are inherently collaborative and require students to take on roles that may or may not give them meaningful access to disciplinary practices \citep{Stump2023, Doucette2020}. \citet{Kalender2021} found that students in labs structured to give them decision-making authority reported stronger experiences of agency and ownership. \citet{DounasFrazer2017} showed that in advanced labs, student ownership and agency were linked to productive struggle and eventual success.

While the links between identity constructs and students' persistence and attitudes are well established, the relationship between identity constructs and epistemic outcomes---specifically, the quality of students' critical thinking as measured by an assessment like the PLIC---has not been studied longitudinally. The cross-lagged panel model (CLPM) design used in this study allows us to examine these directional hypotheses simultaneously within one model.

\section{Methods}

\subsection{Data}

Data were drawn from the international PLIC dataset, which contains responses collected through an automated online administration system \citep{Walsh2019}. The PLIC is deployed via Qualtrics; instructors register through a course information survey and receive a unique link for their course. Students complete the assessment at the beginning and end of the term, and responses are automatically matched using institutional identifiers. We restricted the analysis to students with matched pre- and post-instruction records and complete data on all three PLIC subscales (models, methods, actions). After applying these filters, 5,513 students remained. The dataset spans introductory and beyond-first-year courses at institutions across North America, representing a range of lab curricula from traditional verification-based formats to reformed, skills-focused designs.

For the identity analyses, the dataset additionally included pre- and post-measures of belonging, recognition, self-efficacy, and perceived agency, collected as supplementary items alongside the PLIC. Belonging was measured with items probing students' sense of being accepted and valued in the lab setting. Recognition items asked whether students felt seen as capable by peers and instructors. Self-efficacy items targeted students' confidence in performing specific lab tasks such as data analysis and experimental troubleshooting. Agency items assessed the extent to which students perceived opportunities to make meaningful decisions about their investigations. These constructs are grounded in the physics identity framework developed by \citet{Hazari2010} and operationalized for lab settings by \citet{Kalender2021}. Each construct was measured with a small number of Likert-scale items; while full psychometric validation of these supplementary scales is beyond the scope of this paper, the items draw on established instruments and have been used in prior analyses of the PLIC dataset.

\subsection{Latent Profile Analysis (Study 1)}

We modeled students' response patterns across the three PLIC subscales using latent profile analysis with the \texttt{flexmix} package in R \citep{Grun2008}. The mixture model included Gaussian components for the models and actions subscales and a multinomial component for the methods subscale, reflecting its categorical scoring structure. We estimated solutions with one through four profiles separately at pre and post and evaluated model fit using the Akaike Information Criterion (AIC), Bayesian Information Criterion (BIC), Integrated Classification Likelihood (ICL), posterior classification quality, and entropy-based certainty. For interpretability and direct pre/post comparison, we report the two-profile solution at each timepoint, with profiles relabeled so that Profile~2 consistently represents the higher-performing configuration.

To examine profile transitions over instruction, we linked students' pre and post classifications and computed transition matrices. To assess whether profile composition varied by course type, we conducted chi-square tests of association for the three largest course types in the dataset (Introductory Calculus, Introductory Algebra, and Sophomore-level courses; $N = 5{,}457$ with complete profile assignments at each wave).

\subsection{Cross-Lagged Panel Models (Study 2)}

To examine the longitudinal interplay between PLIC profile membership and identity-related constructs, we estimated cross-lagged panel models (CLPMs) using the \texttt{lavaan} package in R \citep{Rosseel2012}. The model included pre- and post-measures of belonging, recognition, self-efficacy, perceived agency, and Profile~2 posterior membership probability (representing the probability that a student belongs to the higher-knowledge profile). We used maximum likelihood estimation with robust standard errors (MLR) to account for non-normality.

We first estimated a saturated model including all possible cross-lagged paths between the five constructs. We then tested a trimmed model by constraining 10 non-significant cross-lagged paths to zero and compared the two specifications using a chi-square difference test. To assess whether the structural relationships differed by course type, we estimated multigroup CLPMs for the three largest course-type groups (Intro Calculus, $n = 4{,}727$; Intro Algebra, $n = 454$; Sophomore, $n = 276$) and conducted nested Satorra--Bentler difference tests for configural, autoregressive-invariant, and cross-lag-invariant specifications.

\section{Results}

\subsection{Study 1: Latent Profiles of Critical Thinking}

\begin{figure*}[t]
    \centering
    \includegraphics[width=\textwidth]{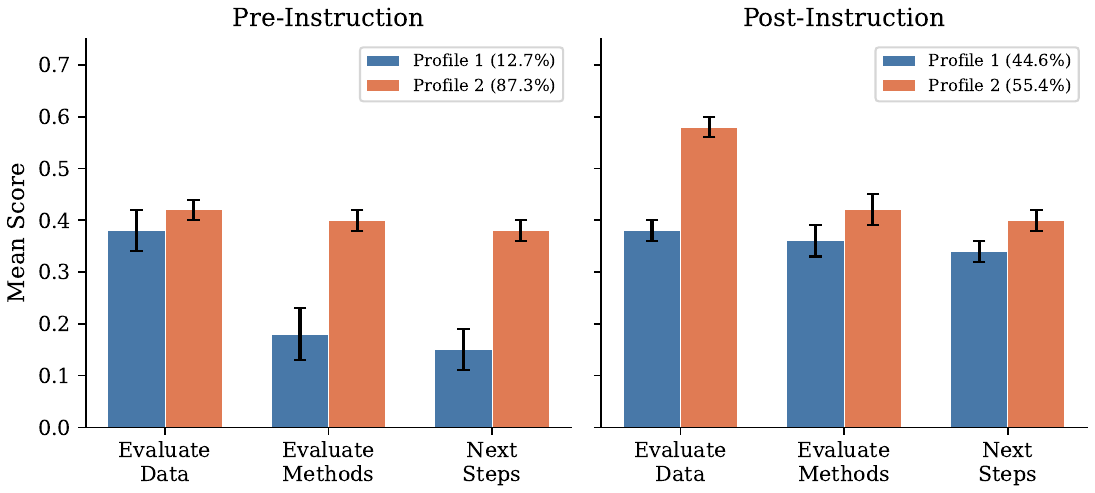}
    \caption{Class-specific mean scores and 95\% confidence intervals on the three PLIC subscales for the two-profile solution at pre-instruction (left) and post-instruction (right). Profile~1 is shown in blue, Profile~2 in orange. Percentages indicate profile membership proportions.}
    \label{fig:profiles}
\end{figure*}

A two-profile solution provided the best balance of fit and interpretability at both pre- and post-instruction timepoints. Solutions with three or four profiles yielded diminishing improvements in information criteria and produced classes with small membership and poor classification certainty; the two-profile solution provided adequate entropy and clear separation between classes. We characterized profiles using class-specific means and 95\% confidence intervals for each of the three PLIC subscales (Fig.~\ref{fig:profiles}).

At pre-test, Profile~1 (12.7\% of students) was characterized by low performance on methods and poor ability to suggest appropriate next steps, with somewhat higher performance on data evaluation. Profile~2 (87.3\% of students) showed relatively uniform performance across all three subscales. At post-test, the profile structure shifted. Profile~1 (44.6\% of students) displayed roughly equal performance across all three subscales at a level comparable to---but slightly lower than---the pre-test Profile~2. Profile~2 (55.4\% of students) was distinguished by strong performance on data evaluation (the models subscale), with more moderate scores on methods and actions.

\begin{figure}[t]
    \centering
    \includegraphics[width=0.95\columnwidth]{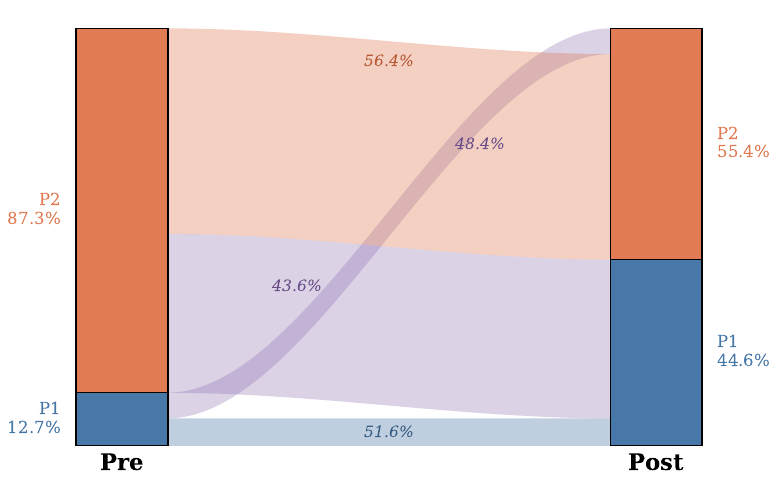}
    \caption{Profile transitions from pre- to post-instruction. Bar heights represent profile proportions. Flow bands connect pre and post profiles; percentages indicate transition rates within each pre-test profile. Purple flows indicate cross-profile transitions.}
    \label{fig:transitions}
\end{figure}

Profile transitions between pre and post were substantial (Fig.~\ref{fig:transitions}). Among students classified in Profile~1 at pre-test, 48.4\% moved to Profile~2 by post-test. Among students classified in Profile~2 at pre-test, 43.6\% transitioned to Profile~1 by post-test. This degree of movement indicates that students' critical thinking profiles are not fixed; instruction reshapes the distribution of students across qualitatively distinct configurations.

Course type was significantly associated with profile composition at both timepoints (pre: $\chi^2(2) = 63.40$, $p < .001$; post: $\chi^2(2) = 55.10$, $p < .001$). Effect sizes, however, were small (Cram\'er's $V = .108$ at pre-test; $.100$ at post-test), indicating that course type explains only a modest share of the variation in profile membership.

\subsection{Study 2: Cross-Lagged Panel Models}

\begin{figure*}[t]
    \centering
    \includegraphics[width=\textwidth]{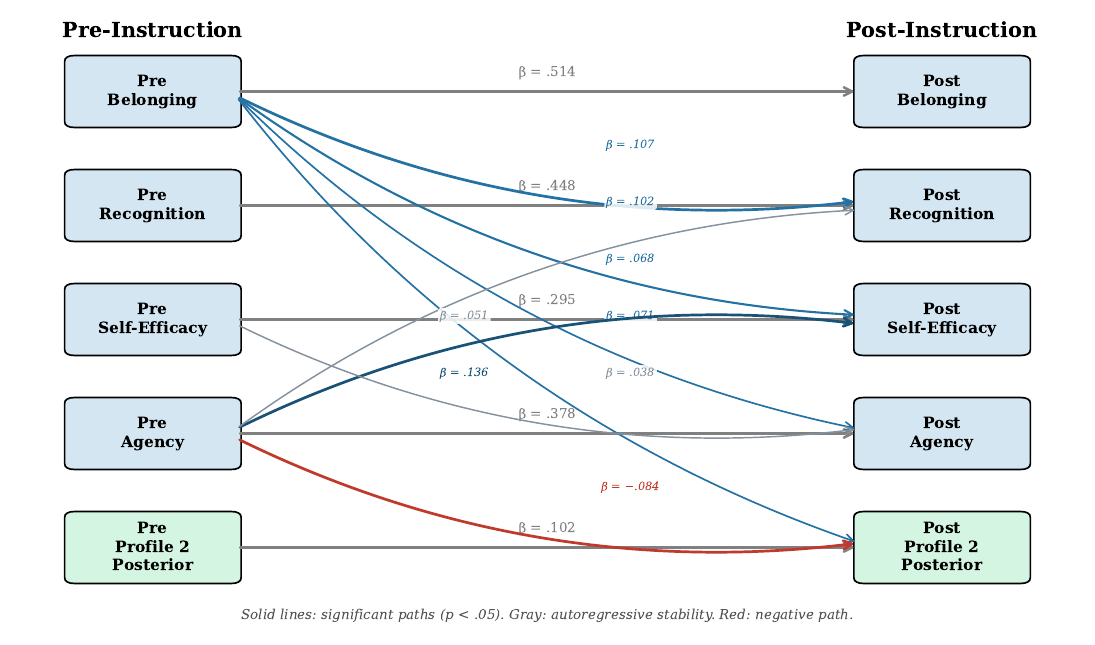}
    \caption{Final cross-lagged panel model. Horizontal gray arrows represent autoregressive stability paths. Colored diagonal arrows represent significant cross-lagged paths ($p < .05$). Blue paths originate from belonging; dark blue from agency; red indicates the negative agency-to-profile path. Path coefficients are standardized.}
    \label{fig:clpm}
\end{figure*}

The trimmed CLPM did not fit significantly worse than the saturated model ($\chi^2(10) = 16.5$, $p = .086$) and showed excellent absolute fit (CFI~$= 1.00$; TLI~$= .998$; RMSEA~$= .011$; SRMR~$= .007$). We retained the trimmed model as the final specification (Fig.~\ref{fig:clpm}).

\textbf{Autoregressive effects.} All autoregressive paths were statistically significant. Standardized stability coefficients ranged from moderate for the affective constructs---belonging ($\beta = .514$), recognition ($\beta = .448$), agency ($\beta = .378$), and self-efficacy ($\beta = .295$)---to low for PLIC profile membership ($\beta = .102$). The comparatively weak stability of profile membership is consistent with the large transitions observed in Study~1 and suggests that critical thinking profiles are more responsive to instruction than identity-related constructs over this timescale.

\textbf{Belonging as an upstream predictor.} Pre-belonging significantly predicted post-recognition ($\beta = .107$, $p < .001$), post-self-efficacy ($\beta = .102$, $p < .001$), post-agency ($\beta = .068$, $p < .001$), and post-Profile~2 membership ($\beta = .071$, $p < .001$). Belonging thus exerted broad prospective influence across both identity-related and epistemic outcomes.

\textbf{Reciprocal agency--self-efficacy dynamics.} Pre-agency predicted post-self-efficacy ($\beta = .136$, $p < .001$) more strongly than pre-self-efficacy predicted post-agency ($\beta = .038$, $p = .026$), consistent with a pattern in which agentic participation precedes gains in efficacy beliefs.

\textbf{Recognition as a downstream construct.} Recognition received input from belonging ($\beta = .107$) and agency ($\beta = .051$) but showed no significant forward-directed paths to other constructs. The path from self-efficacy to recognition was marginal ($\beta = .029$, $p = .074$).

\textbf{Negative agency--profile path.} Pre-agency negatively predicted post-Profile~2 membership ($\beta = -.084$, $p < .001$). Because this result was counterintuitive, we conducted explicit suppression checks: the negative coefficient persisted both with and without belonging in the model, confirming that the effect was not an artifact of collinearity.

\textbf{Multigroup invariance.} Constraining autoregressive paths to equality across course types did not significantly worsen fit (Satorra--Bentler $\Delta\chi^2(10) = 16.34$, $p = .090$). Constraining cross-lagged paths was likewise non-significant at the .05 level ($\Delta\chi^2(40) = 55.23$, $p = .055$). Constraining all regression paths simultaneously reached nominal significance ($\Delta\chi^2(50) = 69.44$, $p = .036$), but follow-up partial-invariance testing showed no significant improvement over the fully constrained model (partial vs.\ all-equal: $\Delta\chi^2(10) = 14.18$, $p = .165$). The single-group model was therefore retained.

\section{Discussion}

\subsection{Critical Thinking as a Multidimensional Profile}

The latent profile analysis demonstrates that students' critical thinking in physics labs is well characterized by distinct, multidimensional profiles rather than a single continuum. The two-profile structure at pre-test separates a small group of students who struggle with methods evaluation and next-step reasoning from a large majority with more uniform---though not necessarily high---performance. After instruction, the profile structure shifts: the higher-performing profile becomes defined by strength in data evaluation (model testing), while the lower-performing profile shows roughly flat performance at a moderate level.

This pattern echoes the three-profile structure identified by \citet{LehtKub2026} in Finnish data, where one profile was similarly characterized by strong data evaluation and another by uniform mid-range performance. That the present analysis, using a much larger and more diverse dataset, identifies a similar structural distinction strengthens confidence in the finding. It also suggests that data evaluation skills may be the dimension most responsive to instruction---a result consistent with the emphasis on model-data comparison in skills-focused lab curricula \citep{Holmes2015, Smith2020}.

The substantial bidirectional movement between profiles over instruction is informative. Nearly half of students in the lower-performing pre-test profile improved to the higher-performing profile by post-test, indicating that lab instruction can produce meaningful gains. At the same time, a comparable fraction of initially higher-performing students moved to the lower-performing profile, raising questions about whether certain instructional approaches fail to sustain incoming competence. The small effect of course type on profile composition suggests that broad curricular categories (e.g., calculus-based vs.\ algebra-based) do not strongly determine which critical thinking configuration students develop. More fine-grained instructional features---such as the extent to which labs emphasize decision making or iterative experimentation \citep{Walsh2022}---may be more consequential.

\subsection{Belonging as a Foundation for Lab Learning}

The CLPM results position belonging as the strongest upstream condition in this instructional window, with prospective effects on recognition, self-efficacy, agency, and epistemic profile membership. This finding is consistent with research demonstrating that social belonging predicts persistence and participation in STEM \citep{Rainey2018, Lewis2017}, and it refines identity-based accounts of physics learning by indicating that belonging may precede recognition in short-term course dynamics \citep{Hazari2010, LockHazari2016}. In the lab context, this makes particular sense: collaborative experimentation makes social inclusion immediately consequential for engagement with disciplinary practices \citep{HolmesWieman2018}. A student who does not feel that she belongs in the lab is unlikely to participate fully in the epistemic work of evaluating data, questioning methods, and proposing next steps.

Recognition, by contrast, functioned primarily as a downstream identity outcome over the pre-post interval. Although recognition is central in science identity theory \citep{CarloneJohnson2007, Hazari2010}, our results suggest that within a single instructional window, recognition is an effect of belonging and agency rather than a driver of them. One interpretation is timescale dependence: recognition may require cumulative social interactions and repeated demonstrations of competence before it exerts forward influence on other constructs \citep{CarloneJohnson2007, Kalender2019}.

\subsection{Agency, Efficacy, and Epistemic Engagement}

The reciprocal but asymmetric relationship between agency and self-efficacy is theoretically coherent with social cognitive theory \citep{Bandura1997}. That the path from agency to later self-efficacy was substantially stronger than the reverse is consistent with the idea that mastery experiences are a primary source of efficacy beliefs. In lab settings, agentic participation---designing procedures, troubleshooting apparatus, making interpretive decisions---can constitute those mastery experiences directly. This interpretation aligns with evidence that ownership-related agency in advanced labs is connected to productive struggle and eventual success \citep{DounasFrazer2017}, and it suggests that lab curricula structured to require meaningful student decision making may be especially effective for building self-efficacy \citep{HolmesWieman2018, Kalender2021}.

The negative cross-lagged path from agency to higher-knowledge profile membership was unexpected and should be interpreted cautiously. The effect survived suppression checks, so it is not simply an artifact of collinearity with belonging. A plausible interpretation is measurement mismatch: self-reported agency may capture initiative and participation without reliably indexing the \emph{epistemic quality} of that engagement. This reading is consistent with work showing that students can be behaviorally active while their epistemological framing remains unproductive for learning \citep{LisingElby2005}, and with resource-based views of epistemology in which productive engagement depends on the activation of appropriate epistemic resources, not merely on participation \citep{HammerElby2003}. Agency, in other words, may not be uniformly beneficial unless it is directed toward model testing, evidence evaluation, and reflective methodological reasoning. Separating behavioral agency from epistemic agency in future measurement models would provide a test of this hypothesis.

\subsection{Future Directions}

Several lines of future research follow from these findings. First, a natural extension of the profile analysis is latent transition analysis (LTA), which would model transitions between profiles probabilistically rather than through post hoc classification. LTA would allow estimation of transition probabilities conditional on instructional features, course type, or demographic variables, providing a more precise picture of who benefits from which kinds of instruction. Second, the negative agency--profile path invites the development of measures that distinguish behavioral agency (initiative, participation, role-taking) from epistemic agency (the quality of reasoning brought to bear on experimental tasks). Third, the present analysis treats course type as a coarse categorical variable; linking profile transitions to specific pedagogical features---such as the extent to which labs emphasize iterative model testing or structured decision making \citep{Walsh2022}---would yield more actionable guidance for instructional design. Finally, extending this work to non-North American contexts is important. The Finnish analyses by \citet{LehtKub2026} and \citet{Pirinen2023} provide a starting point, but systematic cross-national comparison using measurement invariance testing would clarify whether the profile structure identified here generalizes across educational systems.

\subsection{Implications for Practice}

The low autoregressive stability of PLIC profile membership relative to affective constructs is an encouraging result for instructional design: it indicates that epistemic outcomes are comparatively malleable within a single term. The results point toward interventions that strengthen inclusive lab climates and create structured opportunities for student-authored experimental reasoning, while foregrounding the epistemological quality of engagement rather than participation alone \citep{Holmes2015, Walsh2022}. The profile analysis also opens a methodological path for practitioners: by treating critical thinking as a configuration rather than a score, instructors can identify which specific dimensions of critical thinking their students struggle with and tailor instruction accordingly.

\subsection{Limitations}

Several limitations warrant attention. First, the cross-lagged panel model, while allowing examination of directional relationships, cannot establish causation in the absence of experimental manipulation. Second, the identity measures used in this study were brief items embedded in the PLIC administration, not full validated scales; future work with dedicated identity instruments would strengthen the measurement model. Third, our profile analysis used separate solutions at pre and post rather than a longitudinal latent transition analysis, which would model transitions probabilistically rather than through post hoc classification. Fourth, the negative agency path invites further investigation with measures that distinguish behavioral from epistemic agency. Finally, the dataset is dominated by North American institutions, and the generalizability of the profile structure to other educational contexts---while supported by the Finnish findings of \citet{LehtKub2026}---deserves systematic investigation.

\section{Conclusions}

This study provides the first person-centered, multidimensional analysis of PLIC performance in a large, multi-institutional dataset. Students' critical thinking in physics labs is not simply higher or lower; it differs in configuration across the dimensions of data evaluation, methods evaluation, and next-step reasoning. These configurations shift substantially over instruction, with nearly half of students changing profile membership between pre- and post-test.

The cross-lagged panel analysis connects these epistemic profiles to students' identity-related experiences in the lab. Belonging emerges as a foundational condition: it predicts gains in recognition, self-efficacy, agency, and higher-knowledge profile membership. Agency and self-efficacy form a reciprocal but asymmetric loop, with agentic participation driving later efficacy more strongly than the reverse. Recognition, at least over a single instructional window, functions primarily as an outcome rather than a driver. The unexpected negative path from agency to epistemic profile membership raises the possibility that behavioral engagement and epistemic engagement are distinct constructs that must be measured and supported separately.

Taken together, these findings integrate epistemic and affective dimensions of lab learning in a single longitudinal framework and point toward lab designs that combine inclusive climates with structured opportunities for epistemically meaningful student decision making. The person-centered methodology demonstrated here offers a richer account of student learning than composite scores alone and provides a foundation for future work linking profile dynamics to specific instructional features.


\begin{thebibliography}{99}
\small

\bibitem[{AAPT Committee on Laboratories(2014)}]{AAPT2014}
AAPT Committee on Laboratories. (2014). \emph{AAPT recommendations for the undergraduate physics laboratory curriculum}. American Association of Physics Teachers.

\bibitem[{Bandura(1997)}]{Bandura1997}
Bandura, A. (1997). \emph{Self-efficacy: The exercise of control}. Freeman.

\bibitem[{Carlone \& Johnson(2007)}]{CarloneJohnson2007}
Carlone, H.~B., \& Johnson, A. (2007). Understanding the science experiences of successful women of color: Science identity as an analytic lens. \emph{Journal of Research in Science Teaching}, \emph{44}(8), 1187--1218.

\bibitem[{Doucette et~al.(2020)}]{Doucette2020}
Doucette, D., Clark, R., \& Singh, C. (2020). Hermione and the secretary: How gendered task division in introductory physics labs can disrupt equitable learning. \emph{European Journal of Physics}, \emph{41}(3), 035702.

\bibitem[{Douglas et~al.(2014)}]{Douglas2014}
Douglas, K.~A., Yale, M.~S., Bennett, D.~E., Haugan, M.~P., \& Bryan, L.~A. (2014). Evaluation of Colorado Learning Attitudes about Science Survey. \emph{Physical Review Special Topics---Physics Education Research}, \emph{10}(2), 020128.

\bibitem[{Dounas-Frazer et~al.(2017)}]{DounasFrazer2017}
Dounas-Frazer, D.~R., Stanley, J.~T., \& Lewandowski, H.~J. (2017). Student ownership of projects in an upper-division optics laboratory course: A multiple case study of successful experiences. \emph{Physical Review Physics Education Research}, \emph{13}(2), 020136.

\bibitem[{Gr\"un \& Leisch(2008)}]{Grun2008}
Gr\"un, B., \& Leisch, F. (2008). FlexMix version 2: Finite mixtures with concomitant variables and varying and constant parameters. \emph{Journal of Statistical Software}, \emph{28}(4), 1--35.

\bibitem[{Hammer \& Elby(2003)}]{HammerElby2003}
Hammer, D., \& Elby, A. (2003). Tapping epistemological resources for learning physics. \emph{Journal of the Learning Sciences}, \emph{12}(1), 53--90.

\bibitem[{Hazari et~al.(2010)}]{Hazari2010}
Hazari, Z., Sonnert, G., Sadler, P.~M., \& Shanahan, M.-C. (2010). Connecting high school physics experiences, outcome expectations, physics identity, and physics career choice: A gender study. \emph{Journal of Research in Science Teaching}, \emph{47}(8), 978--1003.

\bibitem[{Holmes \& Wieman(2018)}]{HolmesWieman2018}
Holmes, N.~G., \& Wieman, C.~E. (2018). Introductory physics labs: We can do better. \emph{Physics Today}, \emph{71}(1), 38--45.

\bibitem[{Holmes et~al.(2017)}]{Holmes2017}
Holmes, N.~G., Olsen, J., Thomas, J.~L., \& Wieman, C.~E. (2017). Value added or misattributed? A multi-institution study on the educational benefit of labs for reinforcing physics content. \emph{Physical Review Physics Education Research}, \emph{13}(1), 010129.

\bibitem[{Holmes et~al.(2015)}]{Holmes2015}
Holmes, N.~G., Wieman, C.~E., \& Bonn, D.~A. (2015). Teaching critical thinking. \emph{Proceedings of the National Academy of Sciences}, \emph{112}(36), 11199--11204.

\bibitem[{Kalender et~al.(2019)}]{Kalender2019}
Kalender, Z.~Y., Marshman, E., Schunn, C.~D., Nokes-Malach, T.~J., \& Singh, C. (2019). Why female science, technology, engineering, and mathematics majors do not identify with physics: They do not think others see them that way. \emph{Physical Review Physics Education Research}, \emph{15}(2), 020148.

\bibitem[{Kalender et~al.(2020)}]{Kalender2020}
Kalender, Z.~Y., Marshman, E., Schunn, C.~D., Nokes-Malach, T.~J., \& Singh, C. (2020). Damage caused by women's lower self-efficacy on physics learning. \emph{Physical Review Physics Education Research}, \emph{16}(1), 010118.

\bibitem[{Kalender et~al.(2021)}]{Kalender2021}
Kalender, Z.~Y., Stump, E., Hubenig, L., \& Holmes, N.~G. (2021). Restructuring physics labs to cultivate sense of student agency. \emph{Physical Review Physics Education Research}, \emph{17}(2), 020128.

\bibitem[{Lanza \& Rhoades(2013)}]{Lanza2013}
Lanza, S.~T., \& Rhoades, B.~L. (2013). Latent class analysis: An alternative perspective on subgroup analysis in prevention and treatment. \emph{Prevention Science}, \emph{14}(2), 157--168.

\bibitem[{Lehtinen \& Kubsch(2026)}]{LehtKub2026}
Lehtinen, A., \& Kubsch, M. (2026). Characterizing critical thinkers: Latent profiles of Finnish students' critical thinking in the physics lab. \emph{European Journal of Physics}, \emph{47}(2), 025709.

\bibitem[{Lewis et~al.(2017)}]{Lewis2017}
Lewis, K.~L., Stout, J.~G., Finkelstein, N.~D., Pollock, S.~J., Miyake, A., Cohen, G.~L., \& Ito, T.~A. (2017). Fitting in to move forward: Belonging, gender, and persistence in the physical sciences, technology, engineering, and mathematics (pSTEM). \emph{Psychology of Women Quarterly}, \emph{41}(4), 420--436.

\bibitem[{Lising \& Elby(2005)}]{LisingElby2005}
Lising, L., \& Elby, A. (2005). The impact of epistemology on learning: A case study from introductory physics. \emph{American Journal of Physics}, \emph{73}(4), 372--382.

\bibitem[{Lock \& Hazari(2016)}]{LockHazari2016}
Lock, R.~M., \& Hazari, Z. (2016). Discussing underrepresentation as a means to facilitating female students' physics identity development. \emph{Physical Review Physics Education Research}, \emph{12}(2), 020101.

\bibitem[{Pirinen et~al.(2023)}]{Pirinen2023}
Pirinen, P., Lehtinen, A., \& Holmes, N.~G. (2023). Impact of traditional physics lab instruction on students' critical thinking skills in a Finnish context. \emph{European Journal of Physics}, \emph{44}(3), 035702.

\bibitem[{Quinn et~al.(2018)}]{Quinn2018}
Quinn, K.~N., Wieman, C.~E., \& Holmes, N.~G. (2018). Interview validation of the Physics Lab Inventory of Critical Thinking (PLIC). In A. Traxler, Y. Cao, \& S. Wolf (Eds.), \emph{2017 Physics Education Research Conference Proceedings} (pp.\ 324--327). AAPT.

\bibitem[{Rainey et~al.(2018)}]{Rainey2018}
Rainey, K., Dancy, M., Mickelson, R., Stearns, E., \& Moller, S. (2018). Race and gender differences in how sense of belonging influences decisions to major in STEM. \emph{International Journal of STEM Education}, \emph{5}(1), 10.

\bibitem[{Rosseel(2012)}]{Rosseel2012}
Rosseel, Y. (2012). lavaan: An R package for structural equation modeling. \emph{Journal of Statistical Software}, \emph{48}(2), 1--36.

\bibitem[{Smith et~al.(2020)}]{Smith2020}
Smith, E.~M., Stein, M.~M., Walsh, C., \& Holmes, N.~G. (2020). Direct measurement of the impact of teaching experimentation in physics labs. \emph{Physical Review X}, \emph{10}(1), 011029.

\bibitem[{Stump et~al.(2023)}]{Stump2023}
Stump, E.~M., Dew, M., Jeon, S., \& Holmes, N.~G. (2023). Taking on a manager role can support women's physics lab identity development. \emph{Physical Review Physics Education Research}, \emph{19}(1), 010107.

\bibitem[{Walsh et~al.(2022)}]{Walsh2022}
Walsh, C., Lewandowski, H.~J., \& Holmes, N.~G. (2022). Skills-focused lab instruction improves critical thinking skills and experimentation views for all students. \emph{Physical Review Physics Education Research}, \emph{18}(1), 010128.

\bibitem[{Walsh et~al.(2019)}]{Walsh2019}
Walsh, C., Quinn, K.~N., Wieman, C., \& Holmes, N.~G. (2019). Quantifying critical thinking: Development and validation of the physics lab inventory of critical thinking. \emph{Physical Review Physics Education Research}, \emph{15}(1), 010135.

\bibitem[{Wieman \& Holmes(2015)}]{WiemanHolmes2015}
Wieman, C.~E., \& Holmes, N.~G. (2015). Measuring the impact of an instructional laboratory on the learning of introductory physics. \emph{American Journal of Physics}, \emph{83}(11), 972--978.

\end{thebibliography}
\end{document}